\documentclass[prd,aps,floats,twocolumn,nofootinbib]{revtex4-1}
\usepackage{slashed}
\usepackage{mathtools}
\usepackage{amsfonts}
\usepackage{amssymb}
\usepackage{epsfig}
\usepackage{rotating}
\usepackage{url}
\usepackage{times}
\usepackage{color}
\usepackage{bm}
\usepackage{xcolor,colortbl}
\usepackage{hyperref}
\usepackage[alsoload=hep]{siunitx}
\usepackage{enumitem}
\usepackage{fancyhdr}
\usepackage{anyfontsize}
\usepackage{wasysym}
\usepackage{empheq}

\newcommand{\nn}{\nonumber}
\newcommand{\ph}{\phantom}

\newcommand{\be}{\begin{equation}}
\newcommand{\ee}{\end{equation}}
\newcommand{\bea}{\begin{eqnarray}}
\newcommand{\eea}{\end{eqnarray}}

\begin{document}


\title{The real Chern-Simons wave function}
\author{Jo\~{a}o Magueijo}
\email{j.magueijo@imperial.ac.uk}
\affiliation{Theoretical Physics Group, The Blackett Laboratory, Imperial College, Prince Consort Rd., London, SW7 2BZ, United Kingdom}
\date{\today}

\begin{abstract}
We examine the status of the Chern-Simons (or Kodama) state from the point of view 
of a formulation of gravity that uses only real connection and metric variables and a real action. We may package the 
{\it real} connection variables into the complex Self-Dual Ashtekar connection (and will do so to make contact with previous work), 
but that operation is essentially cosmetic and can be undone
at any step or even bypassed altogether. The action will remain the (real) Einstein-Cartan action, forgoing the addition of the usual Holst (or Nieh-Yan) term with 
an imaginary coefficient. It is then found that the constraints are solved by a modification of the Chern-Simons state 
which is a pure phase (in the Lorentzian theory, we stress), the phase containing only the fully gauge-invariant imaginary part of the 
Chern-Simons functional. Thus, the state for the ``real theory'' is non-pathological with regards to the most egregious criticisms facing its ``non-real'' cousin, solving the complex theory. 
A straightforward modification of the real Chern-Simons state is also a solution in quasi-topological 
theories based on the Euler invariant, for which the cosmological constant, $\Lambda$, is dynamical. In that case it is enough to shift the usual factor of $\Lambda$ in the wave function to the inside of the spatial Chern-Simons integral. The trick only  works for the 
quasi-Euler theory with a critical coupling previously identified in the literature. It does not apply to the quasi-Pontryagin theory. 
\end{abstract}

\maketitle

\section{Introduction}
In this paper we re-evaluate the Chern-Simons  wave function (also called the Kodama state~\cite{jackiw,witten,kodama,lee1,lee2}) 
from a different point of view.  We consider a formalism where the gravitational phase space only contains real variables, and the 
action is kept explicitly real. The phase space variables may then be written in terms of the complex Ashtekar Self-Dual connection
to make contact with previous literature and results in Chern-Simons theory, but this ``complexification'' is  essentially cosmetic. It can be undone
at any step or, indeed, be bypassed altogether. No ``reality conditions'' are required, even though our modified solution (which we will call the real Chern-Simons
state) still is more easily written in terms of a complex connection variable. 

Specifically we use the Einstein-Cartan  action as our starting point (later in the paper we will also consider the quasi-topological theories of~\cite{alex1,alex2}). We then consider a Hamiltonian reformulation in terms of a Gauss augmented phase space~\cite{Thomas}, but this is made up
of real triads $E$ (and possibly a time metric variable, if we are not in the ``time gauge'') and real extrinsic curvature $K$ and spatial connection $\Gamma$.
We eschew the canonical transformation (typically assuming zero torsion) that complexifies the theory~\cite{Thomas},
or equivalently, the addition to the Einstein-Cartan action of a Holst (or Nieh-Yan) term with a complex coefficient.  The theory remains real to the bone. 
We then repackage the two real variables $K$ and $\Gamma$ into the Ashtekar Self-Dual connection wherever convenient to connect with the literature.
But, we reiterate, this is not needed. Indeed, later in the paper we will provide a derivation of the real Chern-Simons state in which the Self-Dual variable is never invoked.



We find that the quantum constraints in such a setting are solved, in the connection representation, by a modification of the Chern-Simons state which is always a pure phase
{\it with Lorentzian signature}. Its phase is made up of the usual factors involving the cosmological constant, $\Lambda$,
and the Planck length, $l_P$, and the {\it imaginary} part of the Chern-Simons functional. Postponing detailed definitions (given early in the paper), 
if $Y_{CS}$ is the Chern-Simons functional, then our solution is obtained from the usual Chern-Simons state:
\bea\label{kodcomplex}
\psi_K(A)&=&{\cal N}\exp {\left(\frac{3}{ l_P^2\Lambda} Y_{CS}\right)}
\eea
by the the replacement:
\be\label{modCS}
Y_{CS}\rightarrow i \Im(Y_{CS}).
\ee
If we do not invoke the Self-Dual connection used in the standard definitions of $Y_{CS}$, we obtain this result, in a real formalism, written 
directly in terms of $K$ and $\Gamma$.

Therefore, the state acquires  in the Lorentzian signature the central
property usually ascribed to its version with Euclidean signature~\cite{lee1,lee2,Randono1}: it becomes a pure phase. This is most desirable.
A number of fair criticisms have been levelled against the standard Lorentzian Chern-Simons state (e.g.~\cite{witten}), namely regarding its 
non-normalizability, CPT violating properties (and consequent impossibility of a positive energy property), and lack of gauge invariance under large 
gauge transformations. All of these criticisms hinge on the fact that the state's phase is not purely imaginary, for example 
proportional to $i\Im Y_{CS}$. They evaporate for the our modified Chern-Simons state, just as they do for the Chern-Simons 
state in the Euclidean theory~\cite{Randono1, Randono2}. 

This result is not entirely new: it appears as a limiting case in the work of~\cite{Randono1,Wieland} (the realization that reality 
conditions are not needed in this limiting case might be new, however). So as to connect with this work and with Chern-Simons theory in general~\cite{dunne,Witten},
we can view our approach 
as the result of dissociating two aspects of the Immirzi parameter: the parameter appearing in the definition of the connection, on the one hand, and 
the coefficient of the boundary term added to the action, on the other. We explain this in Section~\ref{2Ims}, stressing that none of this
is needed for a direct derivation (presented later, in Section~\ref{real-formulation}). In Section~\ref{CSreview}  we then review Chern-Simons theory and 
explain how our claim fits in with existing literature.

For the Einstein-Cartan theory, it can be easily proved that our modified Chern-Simons state solves the real Hamiltonian constraint. We first do this
adapting well-known results for the Self-Dual Ashtekar connection (Section~\ref{derivationSD}).  We also show how our
result and derivation reduce to minisuperspace (MSS) results (Section~\ref{MSSreduction}); indeed this paper may be seen 
as a non-perturbative generalization of~\cite{CSHHV,MSS}, which inspired it.
The status of the other constraints is examined in Section~\ref{otherconsts}. The secondary constraints forcing torsion to vanish
do not directly interfere with the topic of this paper (but see~\cite{QuantTorsion}). 
With more relevance here, in Section~\ref{real-formulation}
we show explicitly how the second class reality conditions are not needed in this ``real formalism''. We do this by explicitly rederiving our
result using real variables only, and the uncontaminated Einstein-Cartan action. 

Finally, non-trivially and perhaps surprisingly, we show (Section~\ref{quasi-CS}) that the real Chern-Simons state is also a solution, with minimal adaptation, 
to some quasi-topological theories for which $\Lambda$ is dynamical and varies in space and time~\cite{alex1,alex2}. 
Writing $Y_{CS}=\int {\cal L}_{CS}$, we find a solution by applying to 
real Chern-Simons wave function the replacement
\be
\frac{Y_{CS}}{\Lambda}\rightarrow \int \frac{\cal L_{CS}}{\Lambda} 
\ee
i.e. by passing the varying $\Lambda$  factor inside the Chern-Simons integral. The Hamiltonian and conformal constraints, however, are only consistent
for one class of these theories. Furthermore, even in that case, for the parity-odd branch of the theory the Chern-Simons state is not the only solution. 

We conclude with some comments on the implications of our findings.





\section{The two faces of the Immirzi parameter}\label{2Ims}
Let us consider the Einstein-Cartan action:
\be\label{SEC0}
S_{EC}=\frac{\kappa}{2} \int  \epsilon_{ABCD}\left( e^A e^B R^{CD} -\frac{\Lambda}{6} e^A e^Be^Ce^D\right)  
\ee
where $\kappa=1/(16\pi G_N)$, the indices are $SO(3,1)$ Lorentzian indices, $e^{A}$ are the tetrad 1-forms,  and $R^{AB}=d\Gamma^{AB}+\Gamma^{A}_{\ph{A}C}\Gamma^{CB}$ are the curvature 2-forms of the spin-connection $\Gamma^{A}_{\ph{A}B }$.
We can write this action in terms of the Ashtekar-Barbero $SU(2)$ connection:
\be\label{ashcon}
A^i=\Gamma^i +\gamma_I K^i
\ee
with:
\bea
K^i&=&\Gamma^{0i}\nn\\
\Gamma^i&=&-\frac{1}{2}\epsilon^i_{jk}\Gamma^{jk}\nn
\eea
where $\gamma_I$  the Immirzi parameter and the indices $i,j,k$, etc denote $SU(2)$ components (more on the basis soon). 
We stress that in writing the connection in this format we have {\it not} used the equations of motion
forcing $K^i$ and $\Gamma^i$ to be related to the metric and the torsion. 
If $\gamma_I=i$ the connection is the Self-Dual connection.
Given that the Einstein-Cartan action is real, its complex conjugate (the Anti-Self-Dual connection, $\bar A^i$) will have to appear in any expression of 
$S_{EC}$ in terms of the Self-Dual connection. 

At this stage it may seem that we have complexified the theory, but that is not true, since everything can be rewritten in terms
of the original real variables, $K^i$ and $\Gamma^i$ (as we shall do in Section~\ref{real-formulation}). 
The theory only becomes truly complex if to the Einstein-Cartan action one adds a boundary term, aimed at cancelling off
the terms in $\bar A^i$. This is the Nieh-Yan term:
\bea
S_{NY}&=&\frac{2}{\gamma_{NY}}\int e^Ae^B R_{AB}-T^AT_A .
\eea
(called the Holst term if the torsion is set to zero) 
and its coefficient contains parameter $\gamma_{NY}$,
usually taken to be equal to $\gamma_I$. 
By setting $\gamma_I=\gamma_{NY}=i$ we cancel the Anti-Self-Dual terms that appear in the Einstein-Cartan action, thus 
producing an overall action ($S=S_{EC}+S_{NY}$) that only depends on the Self-Dual connection. 
However, in doing so we have complexified the action and so the theory.

In the usual setting the Immirzi parameter controls {\it both} the definitions of connection and the coefficient of the boundary term,
but  these are conceptually different.
We can work with  $\gamma_I=i$ and at the same time dispense with the Holst/NY term (formally setting $\gamma_{NY}=\infty$). This has the advantage of keeping the action real: it may look as if we have complexified the phase space, but the theory remains real. It is no different from a complex scalar field, for which the action is still real, and which could therefore be rewritten as two real scalar field theories. 

The advantage of this hybrid construction is that it allows us to make use of Chern-Simons theory~\cite{dunne}, namely in solving the (real) Hamiltonian constraint with a  {\it modified} Chern-Simons/Kodama state built from the Ashtekar Self-Dual connection
as well as its Anti-Self-Dual conjugate~\cite{Randono0,Randono1,Randono2,Wieland}. 
But we could equally well dispense with the Ashtekar connection, and derive the final result directly in terms of real
$K^i$ and $\Gamma^i$. For the reader not interested in Chern-Simons theory, we recommend following this alternative route, jumping straight 
into Section~\ref{real-formulation}.

\section{The real Chern-Simons state: background}\label{CSreview}
We now review the standard Chern-Simons state (solving the complex, Self-Dual theory), mainly to define the notation.
We refer the reader to~\cite{dunne} and particularly to~\cite{Wieland} for full details of the results we will be quoting without
proof and using to prove our later results. We then explain how our claim relates to existing literature.

\subsection{Review of the standard Chern-Simons state}
The Chern-Simons functional is usually introduced for topological reasons which will not be relevant here, as we explain straight away.
It is defined as~\cite{dunne,Wieland}:
\be
S_{CS}=k_{CS} \int  {\rm Tr} \left(A dA +\frac{2}{3} A A  A\right)
\ee
where $k_{CS}$ is the ``level'' of the theory. Under large gauge transformations:
\be
S_{CS}\rightarrow S_{CS} -8\pi^2 n
\ee
where $n\in Z$ is the Brouwer degree of the transformation. With the requirement 
that the putative wave function $\psi=e^{iS_{CS}}$ be single valued one  would need the level to be of the form:
\be
k_{CS}=\frac{\rm integer}{4\pi}.
\ee
A similar argument has been used in the context of the Chern-Simons/Kodama state to quantize the cosmological constant (see  \cite{Wieland}).
This line of reasoning is not valid here, because the modified Chern-Simons state we will derive only contains the imaginary part of 
the Chern-Simons functional, which already is invariant under large gauge transformations. Only the real part of $S_{CS}$ transforms 
under large gauge transformations and that will not be part of our proposed wavefunction. 

We will therefore set $k_{CS}=1$ and define the Chern-Simons functional as:
\be
Y_{CS}=\int  {\rm Tr} \left(A dA +\frac{2}{3} A A  A\right).
\ee
Furthermore we will use the basis of  $SU(2)$ generators $t^i=-i\sigma^i/2$, where $\sigma^i$ are Pauli matrices.
This is the basis usually (but not always) used to define the Ashtekar variables, $A^i$, and also the densitized inverse triads $E^i$. 
Hence,  ${\rm Tr } (t^i t^j)=-\delta^{ij}/2$ and  ${\rm Tr } (t^i t^j t^k)=-\epsilon^{ijk}/4$ (and also, for later use, $[t^i,t^j]=\epsilon^{ijk}t^k)$.
Thus, the Chern-Simons functional written in $SU(2)$ components, reads:
\be
Y_{CS}=-\frac{1}{2} \int  A^i dA^i +\frac{1}{3}\epsilon_{ijk} A^i A^j A^k.
\ee
With these definitions it is straightforward~\cite{Wieland} to prove that 
\be\label{id1}
\frac{\delta}{\delta A^i_a(\vec x)}Y_{CS}= - B^a_i(\vec x)
\ee
where $F=dA+AA$ as usual, and the magnetic field is defined from
$F^i_{ab}=\epsilon_{abc}B^{ci }$. The Hamiltonian constraint in the (complex) Self-Dual formalism:
\be\label{Hamconstcomplex}
\epsilon_{abc}  \epsilon^{ij}_{\;\; k} E^a_i E^b_j\left(B^{ck }+\frac{\Lambda}{3}E^{ck}\right)\approx 0
\ee
is then quantum-mechanically solved by the standard Chern-Simons state
presented at the start of this paper, Eq.~\ref{kodcomplex}, in view of (\ref{id1}) and 
since in the quantization diagonalizing the Self-Dual connection we have
\be\label{hatE1}
\hat E^a_i(\vec x)= l_P^2\frac{\delta}{\delta  A^i_a(\vec x) }.
\ee
Here $l_P$ is the reduced Planck length, $l_P=\sqrt{8\pi G_N \hbar}$ (so that 
$l_P^2=\hbar/(2\kappa)$).
Note that in some literature (e.g.~\cite{Thomas}) this is defined as
$l_P=\sqrt{16 \pi G_N \hbar}$ explaining some disparity in factors.


\subsection{Review of previous work on the Chern-Simons state}
Our claimed solution to the real theory can be found in the literature by using $\gamma_I=i$, but $\gamma_{NY}=\infty$.
Then, one arrives at a real version of the complex Hamiltonian 
constraint (\ref{Hamconstcomplex}) which is indeed solved by the standard (\ref{kodcomplex})
with replacement (\ref{modCS}). Taking, for example, Ref.~\cite{Randono1},
setting $\alpha_R=\alpha_L=1/2$ in its Eq.~(4), we see that indeed we do away with the boundary term
($\gamma_{NY}$, called $\beta$ there, becomes infinite). The Chern-Simons state is then modified according to (\ref{modCS}): cf. Eq.~(10) in~\cite{Randono1}. 
Similar results can be found in~\cite{Wieland}, whose notation we follow more closely. 

A point in which 
we differ is in the need to apply ``reality conditions'' to the state in this limit. In this limit the whole theory can be rephrased
in terms of real variables and a real action, as we shall explicitly show in Section~\ref{real-formulation}.

On a related front, we note that we can read off {\it some} of the steps in the derivation below in standard Ashtekar theory.
We refer to~\cite{Thomas} in particular. The point where we should stop ``copying'' results is Section 4.2.1. of ~\cite{Thomas}, 
i.e.: after the ADM space is Gauss augmented, but before a canonical transformation is 
applied (the latter is the equivalent of adding the NY/Holst term). Many of the manipulations complexiifying the theory 
assume the torsion-free condition for the real part of $A^i$. 
We do not want to do this because:
\begin{itemize}
\item We want to consider quasi-topological theories which have an identical Hamiltonian constraint to Einstein-Cartan, but which allow 
for torsion even without spinors, such as the theories of~\cite{alex1,alex2,MZ,MSS}. 
\item Even when the torsion vanishes on-shell we want to leave room for a different quantum mechanical treatment~\cite{QuantTorsion,GenHH}. 
\end{itemize}
We refer to this matter further in Section~\ref{torsionconst}, but stress here that mixing up complexification, reality conditions, and the torsion-free condition
may well have been the undoing of the traditional  Chern-Simons state, Eq.~\ref{kodcomplex}.


\section{The real Hamiltonian constraint in terms of the Self-Dual connection}\label{derivationSD}

We now prove our central claim 
in the most straightforward manner: adapting well-known results for the Self-Dual Ashtekar connection and
CS theory. We also show how our derivation reduces to results found in mini-superspace for Einstein-Cartan theory and 
beyond.


\subsection{Derivation of the state}
Let us consider the Einstein-Cartan action (\ref{SEC0}), and subject it to a 3+1 split in the time gauge, $e^0_a=0$. 
This can be written in terms of the (A)SD connection as:
\be\label{SADM}
S_{EC}=\kappa \int dt \, dx^3 \, \left[2\Im \dot A^i_a E_i^a-(NH + N^a H_a + N_i G^i) \right]
\ee
where $H$, $H_a$ and $G^i$ are the real  Hamiltonian, diffeomorphism and Gauss constraints. As explained, 
the form of the constraints and action can be either found directly, or be lifted from~\cite{Thomas} before complexification and torsion-free conditions 
are imposed, by means of:
\bea
K^i_a&\rightarrow&\Im A^i_a\\
\Gamma^i_a&\rightarrow&\Re A^i_a
\eea
(this can be used as an extra-check). 
Focusing first on the Hamiltonian constraint, we find the real constraint written in terms of Self-Dual quantities:
\bea
H&=  &\epsilon^{ij}_{\;\; k} E^a_i E^b_j\left( \Re F^k_{ab}+\frac{\Lambda}{3}\epsilon_{abc} E^{ck}\right). \label{Hamconstreal}
\eea
instead of the complex (\ref{Hamconstcomplex}). 
From the first term in (\ref{SADM})  we also find that instead of the usual complexified:
\be
\{ A^i_a(\vec x),E^b_j(\vec y)\}=\frac{i}{2\kappa}
\delta^b_a\delta^i_j\delta(\vec x-\vec y)\;  \ee 
(implying
$
\left[ A^i_a(\vec x),E^b_j(\vec y)\right]=- l_P^2
\delta^b_a\delta^i_j\delta(\vec x-\vec y)
$, 
which leads to   (\ref{hatE1}), so that (\ref{Hamconstcomplex}) implies 
(\ref{kodcomplex})), we only have that:
\be\label{PBnonpert}\{\Im A^i_a(\vec x),E^b_j(\vec y)\}=\frac{1}{2\kappa}
\delta^b_a\delta^i_j\delta(\vec x-\vec y)\;  \ee 
implying:
\be\label{Comm}\left[\Im A^i_a(\vec x),E^b_j(\vec y)\right]=i l_P^2
\delta^b_a\delta^i_j\delta(\vec x-\vec y), \ee 
and so:
\be\label{hatE2}
\hat E^a_i(\vec x)=-i l_P^2\frac{\delta}{\delta \Im A^i_a(\vec x) }.
\ee
Since the Hamiltonian constraint instead of (\ref{Hamconstcomplex}) becomes
(\ref{Hamconstreal}), we conclude that  the quantum Hamiltonian equation for the real theory (pure Einstein-Cartan) 
is:
\be\label{quantrealham}
\left ( \Re B^{kc} -\frac{i l_P^2 \Lambda}{3} \frac{\delta}{\delta \Im A^k_c(\vec x) } \right)\psi=0,
\ee
with significant differences with respect to its Self-Dual counterpart.

However, it is easy to adapt the usual derivation of the Chern-Simons state. 
We note that (\ref{id1}) implies:
\be\label{id2}
\frac{\delta}{\delta \Im A^i_a(\vec x)}\Im Y_{CS}= -  \Re B^a_i(\vec x)
\ee
(where it is understood the variation is taken keeping $\Re A^i_a$ fixed). 
The solution to (\ref{quantrealham}) is therefore the modified Chern-Simons state:
\bea\label{kodrealth}
\psi_{CS}(A)&=&{\cal N}\exp {\left(\frac{3i}{ l_P^2\Lambda}\Im Y_{CS}\right)},
\eea
as claimed right at the Introduction.  
To avoid confusion, we shall index this solution with the subscript $CS$ and its complex counterpart,
Eq.~\ref{kodcomplex},  with the subscript $K$.

\subsection{Comparison with mini-superspace}\label{MSSreduction}
It is easy to verify that  the intermediate steps and final result we have obtained  are consistent with the 
equivalent construction in mini-superspace~\cite{CSHHV,MSS}.
For simplicity, consider this exercise for $k=0$ (flat slicing). Then:
\bea
A^i_a&=&\delta^i_a(ib +c)\label{AMSS}\\
E^a_i&=&\delta^a_i a^2.
\eea
The first term in the action (\ref{SADM}) therefore leads to 
\be
S=6\kappa V_c\int dt\, a^2\dot b+...
\ee
 where $V_c$ comes from the spatial integration, and the extra factor of 3 comes
from the degeneracy $\delta^i_a \delta^a_i =3$. Hence, 
\be
\left[\hat b,\hat{ a^2}\right]=\frac{i l_P^2}{3 V_c}\label{com1}
\ee 
is the mini-superspace expression of (\ref{Comm}), consistent with~\cite{MSS}.
Likewise all other steps in the derivation have counterparts in mini-superspace, and if $c\neq0$ we can probe the non-trivial aspects of the
non-complexified construction even in mini-superspace. The final answer found in~\cite{MSS} (and the solution that provides the dual to the Hartle-Hawking and Vilenkin
wave functions~\cite{HH,vil0,Vilenkin}, should we set the torsion to zero~\cite{CSHHV}) is:
\be\label{CSMSS}
\psi(b)={\cal N}\exp {\left(\frac{9iV_c}{ l_P^2\Lambda}\left(\frac{b^3}{3}-bc^2  \right)\right)}. 
\ee
This is just our general result (\ref{kodrealth}) subject to the reduction (\ref{AMSS}). 

Note, however, that we do not need to set the torsion
$c$ to zero by construction ($c$ is Cartan's spiral staircase~\cite{spiral}). 
This is essential, should we wish to treat torsion differently from the usual~\cite{QuantTorsion}, or extend our results to quasi-topological 
theories with torsion~\cite{alex1,alex2,MSS}.

\section{The other constraints}\label{otherconsts}
For completeness, we now investigate the status of the other constraints, which are by and large
irrelevant to this paper. The only matter truly relevant here
concerns the reality conditions, discussed in Section~\ref{real-formulation}.

\subsection{The rotational and diffeomorphism constraints}
As with the standard Chern-Simons state, the real Chern-Simons state is invariant under spatial diffeomorphisms as well as rotations in 
the internal space~\cite{Randono0,Randono1,Randono2,Wieland}. Hence it satisfies the constraints contained in $H_a$ and $G^i$. 

Note that in the real theory these constraints do not need to assume zero torsion. For example, the Gauss constraint obtained
directly from the real Einstein-Cartan theory, but written
in terms of its usual Self-Dual formulation, reads:
\bea
G^i&=&\Im{\cal D}_a E^{ai},
\eea
that is, it is just the ``rotational'' constraint:
\be
\epsilon^{ijk}K^k_aE^a_k=0.
\ee 
We do not need to assume
\be
\Re{\cal D}_a E^{ai}=T^i=0.
\ee
and assumption that mixes the issues of reality and the torsion-free condition. 

\subsection{The torsion constraints}\label{torsionconst}
In Einstein-Cartan theory the torsion-free conditions arise from secondary constraints, which end up forming second class constraints. 
These arise because we have 18 connections ($K^i_a$ and $\Gamma^i_a$) but only 9 metrics $E^a_i$. 

The $\Gamma^i_a$ have a conjugate that is identically zero in the time gauge, where $e^0_a=0$.
In a general gauge the action would start as:
\be\label{SEC}
S_{EC}=\kappa \int dt \, dx^3 \, \left[2\Im \dot A^i_a E_i^a + 2 \Re \dot A^i_a  \epsilon ^{abc}e^0_a e^i_b + ...\right].
\ee
The second term implies that the momentum canonical to $\Re A^i_a=\Gamma^i_a$ is
\be
\Pi_{\Gamma i}^a =\epsilon ^{abc}e^0_a e^i_b,
\ee
and a full covariant treatment of the ensuing second class constraints can be found in~\cite{notimegauge}.
This can be bypassed in the time gauge, where the absence of a $\Pi_{\Gamma i}^a$ may be phrased as
a constraint by rewriting the Einstein-Cartan action as:
\be\label{zeromomconst}
S\rightarrow S+2\kappa \int \dot \Gamma^i_a\Pi_{\Gamma i}^a +\lambda^i_a\Pi_{\Gamma i}^a
\ee
with  $\lambda^i_a$ the corresponding Lagrange multiplier (the normalization $2\kappa$ is irrelevant here, but will be useful
later in the paper). The secondary constraints obtained by evaluating the PB of the Hamiltonian with this constraint then 
force the torsion in  $\Gamma^i$ to be zero, but together with (\ref{zeromomconst})  they are second class constraints.

There are different strategies for dealing with the quantization of systems subject to second class constraints. In this
context usually one solves them classically, by imposing zero torsion before quantizing. However, this is not the only avenue, as we will
demonstrate in~\cite{QuantTorsion}.
Whatever one does, it does not affect the definition of the real Chern-Simons state presented here; it only affects whether the torsion should be left in
the Chern-Simons functional, and conditions upon it placed at a later stater when identifying the physical states
(as we shall do in~\cite{QuantTorsion,GenHH}; or whether torsion should be simply be set
to zero by hand.

\subsection{The reality conditions}\label{real-formulation}
We claimed earlier that we used Self-Dual variables merely to make contact with standard Chern-Simons theory, but that the 
complexification could be bypassed altogether, using only real variables. 
Here we show this explicitly. Backtracking to the Einstein-Cartan action in the time gauge, and writing it in terms of $K^i_a$ and $\Gamma^i_a$
only, we can read off the Poisson bracket:
\be\label{PBreal}\{K ^i_a(\vec x),E^b_j(\vec y)\}=\frac{1}{2\kappa}
\delta^b_a\delta^i_j\delta(\vec x-\vec y)\;  \ee 
implying:
\be\label{Commreal}\left[ K^i_a(\vec x),E^b_j(\vec y)\right]=i l_P^2
\delta^b_a\delta^i_j\delta(\vec x-\vec y), \ee 
and so:
\be\label{hatE3}
\hat E^a_i(\vec x)=-i l_P^2\frac{\delta}{\delta K^i_a(\vec x) }.
\ee
The $\Gamma^i$ have a conjugate momentum forced to be zero, leading to the second class constraints mentioned in the 
previous section, but that is beside the point here. The fact is that we now have a phase space made of explicitly real variables,
as well as a real action. Everything is real by construction. 

This is also true of the Hamiltonian constraint. 
Noting that the 3-curvature in our generator basis is:
\be
^{(3)}\!R^i=d\Gamma^i+\frac{1}{2}\epsilon ^{ijk}\Gamma^j\Gamma^k
\ee
or in spatial components:
\be
^{(3)}\!R^i_{ab}=2\partial_{[a}\Gamma^i_{\; b]}+\frac{1}{2}\epsilon ^{ijk}\Gamma^j_{\; a}\Gamma^k_{\; b}
\ee
and likewise for 
\be
F^i=d A ^i+\frac{1}{2}\epsilon ^{ijk} A ^j A^k
\ee
we can rewrite the (real) Hamiltonian constraint Eq.~(\ref{Hamconstreal}) as:
\be\label{Hamconstreal1}
\epsilon^{ij}_{\;\; k} E^a_i E^b_j\left(^{(3)}\! R^k_{ab}-\frac{1}{2}\epsilon^{klm}K^l_aK^m_b+\frac{\Lambda}{3}\epsilon_{abc} E^{ck}\right) 
=0,
\ee
or its quantum version, with appropriate ordering:
\be\label{Hamconstreal1quant}
\left(^{(3)}\! R^k_{ab}-\frac{1}{2}\epsilon^{klm}K^l_aK^m_b-i\frac{l_P^2\Lambda}{3}\epsilon_{abc} 
\frac{\delta}{\delta K_{ck}}\right) \psi = 0.
\ee
It is then easy to see that a solution is given by:
\begin{widetext}
\bea
\psi&=&{\cal N}\exp{\left[-\frac{3i}{\Lambda l_P^2}\int K^i\,  {}^{(3)}\! R^i  - \epsilon_{ijk}\frac{K^i K^j K^k}{6} \right]}\nn \\
&=&{\cal N}\exp{\left[-\frac{3i}{2\Lambda l_P^2}\int 2 K^i d\Gamma^i  + \epsilon_{ijk}{\left(K^i\Gamma^j\Gamma^k-\frac{K^i K^j K^k}{3}\right)}\right]}\label{realwithreal1}\\
&=&{\cal N}\exp{\left[-\frac{3i}{2\Lambda l_P^2}\int K^i d\Gamma^i +\Gamma^i dK^i + \epsilon_{ijk}{\left(K^i\Gamma^j\Gamma^k-\frac{K^i K^j K^k}{3}\right)}\right]}\label{realwithreal2}
\eea
\end{widetext}
after an integration by parts in the last step. The last expression is nothing but the real Chern-Simons state (\ref{kodrealth}) written in a form invoking explicitly real variables only. Admittedly the expression in terms of complex variables is more elegant, but the point has been made, we hope. 

The integration by parts needed to bring the state obtained directly in terms of real variables (i.e. Eq.~\ref{realwithreal1}) to the form obtained by taking the imaginary part of the complex $Y_{CS}$ (i.e. Eq.~\ref{realwithreal2}) is innocuous here. But it will not be irrelevant in quasi-topological theories, where $\Lambda$ becomes dynamical. This will be an essential clue allowing for the straightforward generalization of the real Chern-Simons state for a class of such theories.


\section{Generalization to quasi-topological theories}\label{quasi-CS}
Perhaps surprisingly, the real Chern-Simons state obtained for the Einstein-Cartan theory generalizes almost trivially to one of the quasi-topological theories in~\cite{alex1,alex2}, as already hinted at in the appendix of~\cite{MSS} (unfortunately with a large number of typos, hopefully 
corrected here). This only applies to the quasi-topological theory
based on the Euler invariant with the critical coupling identified in~\cite{alex1}.
We will keep the Pontryagin term and a general coefficient for the Euler term in part of the analysis, to show this fact.
Except for that one special case, the Hamiltonian and conformal constraints turn out to contradict each other (something we already knew from 
MSS~\cite{MSS}).

In these theories, one adds to the Einstein-Cartan action (\ref{SEC0}) 
the quasi-topological terms:
\bea
S_{QT}&=&\frac{\kappa}{2} (
S_{Eul}+ S_{Pont})\\
S_{Eul}&=&-\frac{3\theta}{2}\int \frac{1}{\Lambda} \epsilon_{ABCD}R^{AB}R^{CD} ,\label{euler}\\
S_{Pont}&=&-\frac{3}{\gamma_P}\int \frac{1}{\Lambda} R^{AB}R_{AB}.
\eea
The critical coefficient for the Euler theory is obtained with $\theta=1$, as explained in~\cite{alex1,alex2}.
Arguments were given in~\cite{MZ} for relating the coefficient of the Pontryagin term, $\gamma_P$, to the Immirzi parameter
(specifically its aspect denoted by $\gamma_{NY}$ here).
These arguments are merely motivational, so we shall not relate $\gamma_P$ with either $\gamma_I$ 
or $\gamma_{NY}$.

In such theories $\Lambda$ becomes a dynamical variable, so that in this Section  $\Lambda\equiv \Lambda(x)$
(in contrast with the rest of this paper where $\Lambda\equiv \Lambda_0$). 
Our claim is that for the critical Euler theory, a solution to the Hamiltonian and conformal constraints can be obtained 
directly from Eq.(\ref{kodrealth}) by replacing 
$Y_{CS}$ by:
\be\label{YLambda}
Y_{\Lambda}=\int \frac{{\cal L}_{CS}}{\Lambda}
\ee
with:
\bea
{\cal L}_{CS}&=& {\rm Tr} \left(A dA +\frac{2}{3} A A  A\right)\nn\\
&=&
-\frac{1}{2} \left( A^i dA^i +\frac{1}{3}\epsilon_{ijk} A^i A^j A^k\right).\label{CalLCS}
\eea
That is, 
it is enough to move the space-time varying $\Lambda$ in (\ref{kodrealth}) inside the integral $Y_{CS}$. 
The claimed solution is therefore: 
\bea\label{kodrealtha}
\psi_\Lambda (A)&=&{\cal N}\exp {\left(\frac{3i}{l_P^2}\Im Y_{\Lambda}\right)}.
\eea
As we have already found in~\cite{MSS} for mini-superspace, this solution is only a particular solution, for the parity-odd branch of the theory.
A more general solution accepts a non-constant amplitude ${\cal N}$ with a particular functional dependence. This issue, as well as the full algebra
of constraints for these theories is deferred to~\cite{MZ1}. Suffice it to say here that it is possible to implement the other constraints in a different form,
as can be done in the Einstein-Cartan theory for the torsion~\cite{QuantTorsion}.

To find the general form of the constraints, we note that, since:
\be
R^i R^i=\frac{1}{2}{\left( R^{AB}R_{AB}  -\frac{i}{2}
 \epsilon_{ABCD}R^{AB}R^{CD}\right)},
\ee
the quasi-topological terms 
can be written as:
\be
S_{QT}=
-\kappa \int \frac{1}{\Lambda} \Re (\zeta R^i R^i) 
\ee
with:
\be\label{zeta}
\zeta=3\left(\frac{1}{\gamma_P} +i\theta \right).
\ee
But given the standard result:
\be
{\rm Tr} RR=d{\cal L}_{CS}
\ee
or in components:
\be\label{potential}
R^i R^i=d\left(A^idA^i+\frac{1}{3}\epsilon_{ijk} A^i A^j A^k\right)
=-2d{\cal L}_{CS}
\ee
we have:
\be\label{SQT1}
S_{QT}=
2\kappa \int \frac{1}{\Lambda}d\Re( \zeta {\cal L}_{CS}).
\ee
Subjecting this action to a 3+1 split generates 3 types of terms, depending on where in the integrand 
the time index is. We proceed to examine them.

\subsection{The conformal constraint}
The first type of term is of the form:
\be\label{SQT1T1}
S=-2\kappa \int dt d^3x\,  \dot \phi \Re( \zeta {\cal L}_{CS}) +...
\ee
where we set $\phi=1/\Lambda$, and integrated by parts in time. This suggests a PB between $\phi$ and its conjugate momentum:
\be
\{\phi({\bf x}),\Pi({\bf y})\}=-\frac{1}{2\kappa}\delta({\bf x}-{\bf y})
\ee
subject to the primary constraint:
\be\label{confconst}
{\cal V}=\Pi - \Re( \zeta {\cal L}_{CS})\approx 0.
\ee
Eq.~\ref{confconst} is the generalization beyond mini-superspace of the conformal constraint found in~\cite{MZ}. It differs from its
MSS version in that it must be enforced point by point. 

Elsewhere~\cite{MZ1} we will show that the total system of constraints is first class in the case relevant to this paper. 
Therefore, quantum mechanically we can promote the PB to:
\be
\left[\hat \phi({\bf x}),\hat \Pi({\bf y})  \right]=-il_P^2 \delta({\bf x}-{\bf y})
\ee
so that in the $\phi$ representation we have:
\be
\hat \Pi({\bf x}) =il_P^2\frac{\delta}{\delta\phi({\bf x})}.
\ee
The quantum conformal constraint equation
\be
\hat {\cal V}\psi=\left(il_P^2\frac{\delta}{\delta\phi({\bf x})} - \Re( \zeta {\cal L}_{CS})\right)\psi= 0
\ee
is solved (point by point) by:
\bea\label{realKod}
\psi&=&{\cal N}\exp{\left(-\frac{i}{ l_P^2 }\Re \zeta Y_{\Lambda }\right)}\nonumber\\
&=&{\cal N}\exp{\left[\frac{3i}{ l_P^2}
\left(\theta \Im Y_{\Lambda} -\frac{1}{\gamma_P}\Re Y_{\Lambda}\right)\right]}.
\eea
At once we note that we only get a form that generalizes the standard Chern-Simons state (i.e. Eq.~\ref{kodrealtha})
if $\theta=1$ and $\gamma_P=\infty$. We now show that this is the only case where the solutions of the conformal
and Hamiltonian constraints coincide.



\subsection{The Hamiltonian constraint}
Beside the term highlighted in Eq.~(\ref{SQT1T1}), the action (\ref{SQT1})  contains 
two other types of term, bearing Eq.~(\ref{CalLCS}) in mind.  Terms in $A^i_0$ generate a new constrain, since $A^i_0$ does not have
a time derivative anywhere, and so can be regarded as a Lagrange multiplier. 
This new constraint, together with the other constraints and their algebra, will be studied in~\cite{MZ1}. In addition there are terms in $\dot A^i_a$, 
leading to the contribution:
\bea\label{SQT1T2}
S&=& \kappa \int dt d^3x\, \epsilon^{abc}  (\partial_a \phi) \Re( \zeta A^i_b \dot A^i_c) +...\nn\\
&=&   \kappa \int dt d^3x\, \epsilon^{abc}  (\partial_a \phi) [-3\theta (K^i_b\dot\Gamma^i_c + \Gamma^i_b\dot K^i_c )+\nn\\
&&+\frac{3}{\gamma_P}(-K^i_b\dot K^i_c+\Gamma^i_b\dot \Gamma^i_c)]   + ...
\eea
The term in $\dot \Gamma$ implies that no longer the momentum conjugate to $\Gamma$ is set to zero in the time gauge
by a primary constraint, as was the case in Section~\ref{torsionconst}. Instead, we have:
\bea
\{\Gamma^i_a(\vec x),\Pi_{\Gamma j}^b (\vec y)\}&=&\frac{1}{2\kappa}
\delta^b_a\delta^i_j\delta(\vec x-\vec y)\nn\\
\Pi^{a i}_{\Gamma}&\approx & \frac{3}{2}\epsilon ^{abc}\left(\theta K^i_b - \frac{1}{\gamma_P}\Gamma^i_b\right)\partial_c \phi.
\eea
The conditions upon the torsion that follow will therefore be modified, a matter we 
examine thoghroughly in~\cite{MZ1}.  Again one may choose to impose the torsion conditions on the classical theory and then quantize; or, rather, to impose them on the wavefunction only, a matter we highlight here, but which is beyond the scope of 
this paper~\cite{QuantTorsion}.

The term in $\dot K$  affects instead Eqs.~(\ref{PBnonpert}) and (\ref{PBreal}), which 
retain their form, for example:
\be\label{PBreal1}\{K ^i_a(\vec x),\Pi^b_j(\vec y)\}=\frac{1}{2\kappa}
\delta^b_a\delta^i_j\delta(\vec x-\vec y)\;  \ee 
but with:
\be\label{NewPi}
\Pi^a_i= E^a_i+\frac{3}{2}\epsilon ^{abc}\left(\theta \Gamma^i_b+\frac{1}{\gamma_P}K^i_b\right)\partial_c \phi.
\ee
In the representation diagonalizing the connection we therefore now have:
\be\label{NewE}
\hat E^a_i=-i l_P^2\frac{\delta}{\delta K^i_a  } -\frac{3}{2}\epsilon ^{abc}
\left(\theta \Gamma^i_b+\frac{1}{\gamma_P}K^i_b\right)
\partial_c \phi,
\ee
and it will be this representation that we should use to implement the Hamiltonian and other constraints.

At this point we note that once we account for the 3 types of term into which (\ref{SQT1}) can be expanded, 
nothing is left that can alter the other constraints. In particular the Hamiltonian constraint in these theories retains the Einstein-Cartan form
(Eqs.~(\ref{Hamconstreal}) or (\ref{Hamconstreal1})), with the trivial modification $\Lambda\rightarrow 1/\phi(x)$, i.e.: Lambda becomes a field.  However, 
this is only true classically: the quantum mechanical expression of the constraint is modified, because the 
form of the operator $\hat E$  changes from (\ref{hatE2}) or (\ref{hatE3})  to (\ref{NewE}). 
The quantum Hamiltonian constraint, with suitable ordering, is therefore:
\begin{widetext}
\be\label{quantumhamLambda}
\left[\Re F^k_{ab}+ \frac{1}{3\phi}\epsilon_{abc}
\left(i l_P^2\frac{\delta}{\delta K^k_c  } 
+  \frac{3}{2}\epsilon ^{abc}
\left(\theta \Gamma^i_b+\frac{1}{\gamma_P}K^i_b\right)
\partial_c \phi
\right)
\right]\psi
=0.
\ee
\end{widetext}

We now note that the algebraic manipulations leading to (\ref{id2}) carry over trivially into our calculation by replacing $Y_{CS}/\Lambda_0$
with $Y_\Lambda$, because $\phi$ appears in them as an irrelevant multiplicative factor. The exception is when we have to integrate by parts: what usually produces a boundary term that can be discarded now produces terms in $d\phi$. 
Thus, inserting $\psi=\psi_\Lambda$ (defined in Eq.~(\ref{kodrealtha})) in Eq.~(\ref{quantumhamLambda}), we find that its first two terms
cancel as usual, if we ignore the boundary term.  However, the usually innocuous integration by parts now leads to an extra term:
\be
\left[\Re F^k_{ab}+\frac{1}{3\phi}\epsilon_{abc}
\left(-i l_P^2\frac{\delta}{\delta K^k_c  }\right)
\right]\psi_\Lambda=
\left(\frac{3}{2}\epsilon ^{cde}\Gamma^i_d\partial_e \phi\right)\psi_\Lambda.
\ee
This term cancels the extra terms in the Hamiltonian equation (\ref{quantumhamLambda}) if
$\theta=1$ and $\gamma_P=\infty$, completing the proof of our claim.


\section{Conclusions}
We conclude with a digression on the origin of our result, followed by a discussion of some of  its implications.

Complexification, as in the Self-Dual formalism of General Relativity, is a wonderful tool but presents some dangers. 
Foremost, in complexifying the theory one often folds in the assumption that the connection is torsion-free. 
Indeed, most manipulations leading to the Self-Dual formulation~\cite{Thomas} without spinors assume that the connection is torsion-free (see~\cite{Mercuri,Calcagni} 
for the situation with spinors and not only). This may obscure matters, particularly when we do not want to assume zero torsion 
off-shell~\cite{QuantTorsion}, or in extensions to Einstein-Cartan theory~\cite{alex1,alex2,MSS}  
where we simply cannot assume it, even though the Hamiltonian constraint is unmodified. 
This, we would argue, is behind the apparent pathologies of the Chern-Simons state in Self-Dual theory. None of them are present
in an explicitly real formalism where no assumptions on torsion (folded into the complexification and the reality conditions)
have yet been made at the point of quantization.

For example, by taking the full complex Hamiltonian constraint in mini-superspace (and not just its real part),
we require in addition to the real part condition that:
\be
\epsilon^{ij}_{\;\; k} E^a_i E^b_j  \Im F^k_{ab}\propto \Im (c+ib)^2=2bc=0,
\ee
where, we recall, $b$ and $c$ are the connection components defined in (\ref{AMSS}) (the latter the parity-violating Cartan spiral staircase~\cite{spiral}). 
This is overkill on various levels. Firstly, it cannot be assumed for quasi-topological theories, blocking the calculations in Section~\ref{quasi-CS}.
Indeed, even for the quasi-Euler theory with critical coupling we have  $c=c_0\neq 0$ in the absence of matter (and more complicated solutions 
in the presence of matter). 

Secondly, we may not want to assume zero torsion before quantization even in Einstein-Cartan theory~\cite{QuantTorsion}. We may want to define a 
real Chern-Simons state with unconstrained torsion, and then build wave packets indexed by the torsion, around zero torsion, or around any $c_0$  in the quasi-Euler
theory. The reason is that this illuminates the meaning of normalizability for the wave function. As suggested in~\cite{Randono1,Randono2},
the non-pathological real Chern-Simons state is delta-function normalizable, but what does this mean? With respect to what? A possible answer
is: with respect to torsion~\cite{QuantTorsion}. The wave-packets in torsion space are then normalizable in the conventional sense. 

This sheds light on the issue of the normalization and interpretation of the wave function of the Universe (in particular of the 
Hartle-Hawking variety~\cite{HH}, although an adaptation of the construction may work for the Vilenkin one~\cite{HH,vil0,Vilenkin,CSHHV}). 
The real Chern-Simons state with zero-torsion, when reduced to mini-superspace is the Fourier dual of the Hartle-Hawking wave function, with a real integration 
contour~\cite{CSHHV}. Hence, normalizability issues should be identical for both, since they do not depend on the representation. We can therefore
import the construction proposed above (wave-packets in torsion space) into the metric formulation. This leads to a regular
Hartle-Hawking ``beam''~\cite{QuantTorsion}. None of this would be possible with the non-real Chern-Simons state, which would acquire the pathological extra factor:
\be
\psi_{CS} \rightarrow \psi_{CS} \exp {\left(\frac{3V_c}{ l_P^2\Lambda}\left(c^3-3b^2c  \right)\right)},
\ee
applied to (\ref{CSMSS}). 

Another implication of our result is that, since the  Chern-Simons state is a non-perturbative construction, 
and its pure phase property is general too, it should be possible to define generic metric representations by standard Fourier Transform.
What are its metric duals beyond mini-superspace, for example for anisotropic and inhomogeneous models, black holes
or gravity waves? These can be seen as generalizations of the Hartle-Hawking wave function and a preliminary study will be 
presented in~\cite{GenHH}.

We close with a couple of speculations. Could a closer examination of our proposal show that it is valid,  
not for standard Einstein-Cartan theory, but for a variation on the theory
which breaks local Lorentz invariance? It could be that the way we dealt with the constraints fixes the frame in which the 3+1 
decomposition was performed, so that a 3+1=4 cannot be reinstated (in vague analogy with the Horava-Lifshitz construction~\cite{Horava}). 
Care was taken to avoid this possibility (and this author has found no evidence
for it), but it is possible that hiding somewhere is the peg for a preferred Lorentz frame. The same concern may be voiced about
$SU(2)$ symmetry breaking. We can also speculate about  some other points of overlap with the concepts of holography~\cite{hololee,hololee1}
or new approaches to Chern-Simons theory~\cite{Witten}.


\section*{Acknowledgements}
This paper grew out of an earlier project~\cite{MSS} and was finished in parallel with 2 projects~\cite{MZ1,GenHH}, in collaboration
with Stephon Alexander, Gabriel Herczeg, Simone Speziale and Tom Zlosnik, to whom I am very grateful. 
In addition I thank Lee Smolin and Thomas Thiemann for advice and discussions. 
This work was supported by the STFC Consolidated Grant ST/L00044X/1.


\end{document}